\documentclass[12pt,preprint]{aastex}

\shorttitle{NGC 1399 Nuclear Activity } 

  \newcommand{\mathfont}[1]{\ifmmode{#1}\else{$#1$}\fi}

  \newcommand\kms{\ifmmode{{\rm km\ s}^{-1}}\else{${\rm km\ s}^{-1}$}\fi}

  \newcommand\angstrom{\alwaysmath{\,{\rm A^{\kern -0.53em {\tenrm ^\circ}}}}}

\begin{document}

\title{UV/Optical Nuclear Activity in the gE Galaxy NGC 1399  } 



\author{Robert W.\  O'Connell\altaffilmark{1}, Jodie R. Martin\altaffilmark{1},
Jeffrey D. Crane\altaffilmark{1} }

\author{David Burstein\altaffilmark{2}, Ralph C. Bohlin\altaffilmark{3},
Wayne B. Landsman\altaffilmark{4} }  

\author{Ian Freedman\altaffilmark{1} and  Robert T. Rood\altaffilmark{1}}

\altaffiltext{1}{Astronomy Department, University
of Virginia, P.O.\ Box 3818, Charlottesville, VA 22903-0818}

\altaffiltext{2}{Dept. of Astronomy \& Physics, Arizona State University,
Tempe, AZ 85287-1504}

\altaffiltext{3}{Space Telescope Science Institute, 3700 San Martin
Dr., Baltimore, MD 21218}

\altaffiltext{4}{SSAI, NASA--Goddard Space Flight Center, Code 681,
Greenbelt, MD 20771}


\begin{abstract}

Using HST/STIS, we have detected far-ultraviolet nuclear activity in
the giant elliptical galaxy NGC 1399, the central and brightest galaxy
in the Fornax I cluster.  The source reached a maximum observed far-UV
luminosity of $\sim 1.2 \times 10^{39}$ ergs s$^{-1}$ in January 1999.
It was detectable in earlier HST archival images in 1996 (B band) but
not in 1991 (V band) or 1993 (UV).  It faded by a factor of $\sim
4\times$ by mid-2000.  The source is almost certainly associated with
the low luminosity AGN responsible for the radio emission in NGC
1399.  The properties of the outburst are remarkably similar to the
UV-bright nuclear transient discovered earlier in NGC 4552 by Renzini
et al.\ (1995).  The source is much fainter than expected from its
Bondi accretion rate (estimated from {\it Chandra} high resolution
X-ray images), even in the context of ``radiatively inefficient
accretion flow'' models, and its variability also appears inconsistent
with such models.  High spatial resolution UV monitoring is a valuable
means to study activity in nearby LLAGNs.

\end{abstract}



\section{Introduction}



There is now convincing kinematic evidence that most large elliptical
and spiral galaxies contain supermassive nuclear black holes (SMBHs)
with masses proportional to the masses of their spheroidal components
(e.g.\ Magorrian et al.\ 1998, Ferrarese \& Merritt 2000,  Gebhardt et al.\
2000; and reviews by Kormendy \& Richstone 1995 and Ho 1998).  Only a
small fraction of these exhibit the prodigious central energy releases
traditionally associated with active galactic nuclei.  Instead, most
are ``low-luminosity active galactic nuclei'' (LLAGN's), with
bolometric luminosities less than 1\% of their Eddington luminosities
(e.g.\ Ho 2003).  

In principle, the shape and variability of the multiband
electromagnetic spectrum (radio through X-ray) of LLAGNs contains a
great deal of information about energy conversion mechanisms in the
vicinity of the SMBH and the processes which feed the activity (see \S
4).  Only recently, however, have observational facilities improved to
the point that LLAGN can be explored at levels below $L
\sim 10^{40}$ ergs s$^{-1}$ at UV/optical wavelengths in nearby
galaxies.  With small aperture spectroscopy, Ho and his
colleagues find signatures of nuclear activity in the form of optical
emission lines (e.g.\ LINERS) in about half of all nearby galaxies
(Ho, Filippenko \& Sargent 1997).  There are remarkably large
ranges in the shape of the LLAGN spectra and in their luminosity
relative to the estimated SMBH mass (Ho 2002).  The faintest LLAGNs
in the emission-line samples have $\lambda L_{\lambda} \sim 10^{38-39}$ ergs
s$^{-1}$ at optical wavelengths.  

In addition to their importance in spectral shape tests of AGN mechanisms, vacuum
ultraviolet ($\lambda \sim 1100$-$3000$ \AA) observations with the {\it Hubble}
Space Telescope (HST) can push down the detection threshold for
LLAGNs for two reasons.  First, the high spatial resolution of the HST
(with a point-spread-function area $\sim 200 \times$ smaller than at
ground-based observatories) allows better isolation of faint point
sources in the galaxy cores.  Second, the flux contrast between the
flat energy distribution  of an AGN
and the steep spectral background of a typical galaxy core with an older
stellar population dramatically improves in the UV, by factors up to
100$\times$ over that in the V-band (e.g.\ O'Connell 1999). 
HST surveys (Maoz et al.\ 1995, Barth et al.\ 1998) for UV-bright
nuclear sources show nuclear peaks in excess of the interpolated
stellar background in about 15\% of the cases, but not all of these are
AGN's.  Some are clearly young star clusters or other extended sources
(e.g.\ NGC 2681, Cappellari et al.\ 2001).  Source variability is an
important means of distinguishing true AGN's from these other phenomena. 

One of the most intriguing instances of unusual LLAGN activity 
was the serendipitous discovery by Renzini et
al.\ (1995) of a transient UV point source in the center of the Virgo
Cluster giant elliptical NGC 4552 on UV exposures with the HST/Faint
Object Camera.  The source was detected at three epochs (1991, 1993,
and 1997), during which it brightened by a factor of 4.5$\times$, then
faded.  Although NGC 4552 was known to host a variable compact radio
source (e.g.\ Wrobel \& Heeschen 1984), there had been no earlier
evidence for activity in the UV/optical bands, and the galaxy had not been
classified as a LINER.  Spectroscopy by Cappellari et al.\ (1999)
showed that the source was indeed a faint LINER/Seyfert nucleus with a
broad H$\alpha$ emission line luminosity of $\sim 5.6 \times 10^{37}$
ergs s$^{-1}$.  The outburst's UV luminosity was $\lambda
L_{\lambda}({\rm UV}) \sim 7.5 \times 10^{38}$ ergs s$^{-1}$, and its
estimated bolometric luminosity was $L_{\rm bol} \sim 3 \times 10^5\,
{\rm L}_{\odot}$, comparable to the faintest optical-band detections.
ROSAT detected a compact nuclear X-ray source in NGC 4552 with a
luminosity of $L_{\rm X} \sim 5 \times 10^{40}$ ergs s$^{-1}$ (Brown
\& Bregman 1998; Beuing et al.\ 1998; Schlegel, Petre, \& Loewenstein
1998).  Nuclear UV variability has also been detected in the LINER
nucleus of the Sb galaxy NGC 4579 (Barth et al.\ 1996) and in the
well-known AGN and jet of the gE galaxy M87 (Perlman et al.\ 2003).

In this paper we report the discovery on a spectrum taken with the
{\it Hubble} Space Telescope Imaging Spectrograph (STIS) of a
UV-bright active nucleus in NGC 1399, another outwardly normal gE galaxy.
NGC 1399 is the central and brightest ($M_{\rm B} =
-21.3$) object in the Fornax I cluster of galaxies.  It hosts a small
cluster X-ray bright cooling flow with a formal mass deposition rate of
$\sim 2\, {\rm M}_{\odot}$ yr$^{-1}$ (Rangarajan et al.\ 1995).
Extensive studies have been made of its large globular cluster system
(e.g.\ Dirsch et al.\ 2003 and references therein).  Unlike M87 in the
Virgo Cluster, which it otherwise resembles, NGC 1399 had exhibited
only modest signs of nuclear activity, in the form of a weak radio
source.  It is not classified as a LINER. Its radio and X-ray
properties are discussed further in $\S 4$.  We adopt a distance of
20.3 Mpc for the Fornax cluster based on surface brightness fluctuation
measurements by Jerjen (2003).

We describe our HST spectroscopy in $\S 2$.  In $\S 3$ we combine
that with a reanalysis of other HST observations taken from the HST
archive to describe the properties of the nuclear point source.  In
$\S 4$ and $\S 5$ we interpret the results in the context of other
data on NGC 1399 and recent models for nuclear activity.  

\section{Spectroscopic Observations}

We observed the center of NGC 1399 with the HST/STIS
and the Far-UV MAMA detector on 19 January
1999 as part of HST GO Program 7438.  We used the low-resolution G140L
grating, which  covers 1150-1700 \AA.  Because we were interested in
determining the spectrum of the diffuse UV stellar radiation in the
galaxy, we used a 2\arcsec\ wide slit, which is not well suited for
detection of point sources.  This yields a mean effective spectral
resolution of only 48 \AA\ but provides full STIS resolution (pixel
size 0.024\arcsec or 2.4 pc) in the spatial direction.  The slit was
oriented in PA 30$^{\circ}$.  It was centered on the nucleus and
extended $\pm 12\arcsec$ from the nucleus.  The total exposure time
was 11239 sec, broken into four exposures.  

A week earlier (10 January 1999) we had also obtained a complementary
optical-band spectrum (5402 sec exposure) with the STIS/CCD detector
in the same slit position but with a narrower 0.2\arcsec\ wide slit.
The spatial scale per pixel is 0.051\arcsec.  With grating G430L, the
spectrum covers 2900-5650 \AA\ at a resolution of 2.7 \AA\ per pixel. 

The exposures in each camera were reduced with the standard STIS
pipeline of the Space Telescope Data Analysis System and co-added
together to produce two 2-dimensional images in (wavelength, radius)
coordinates.  We applied a time-dependent correction for dark count
background, but otherwise reduction of the FUV data was routine.
However, the optical-band spectrum required special treatment for
removal of numerous cosmic rays.

\section{Results: A Nuclear Point Source}

NGC 1399 was already known to be unusually bright for an elliptical
galaxy in the far-ultraviolet.  It has the highest ratio of FUV to
V-band light (i.e.\ the strongest ``UV-upturn"
component) among nearby gE's (Burstein et al.\ 1988) and has an
exceptionally high central UV surface brightness, with an average
$\mu_{\lambda}(1500\, {\rm \AA})
\sim 19.2$ mag arcsec$^{-2}$, within $r = 3\arcsec$ (O'Connell et al.\
1992).  This light is produced predominantly by hot ``extreme
horizontal branch'' stars in the normal old stellar population of gE
galaxies (O'Connell 1999 and references therein).  Brown et al.\ (2002)
used the Far Ultraviolet Spectroscopic Explorer to measure stellar absorption
lines in the 900-1200 \AA\ region of the hot component; they obtained
abundances for N, Si, and C consistent with the expected overall high
metallic abundances in a gE galaxy coupled with gravitational diffusion
in the hot stellar atmospheres.  

A spatial profile of our STIS FUV spectral image covering the range 1340-1710
\AA\ (which avoids the strong airglow lines at shorter wavelengths) is
shown in Figure 1.  The plot shows the expected, smoothly extended UV
continuum profile.  But there is a distinct, pointlike source at the
core of the profile.  The source lies within $\sim 0.04\arcsec$ (4 pc)
of the photometric center of the galaxy as defined by the symmetry of
the background UV stellar light.  Because NGC 1399 has the broad
core-like central light profile characteristic of luminous E galaxies,
this source is not related to the (stellar) light spikes found in lower
luminosity galaxies with power law profiles (Stiavelli, Moller, \&
Zeilinger 1993; Lauer et al.\ 1995; Faber et al.\ 1997).

We used the profile shown in the figure to estimate the mean flux of
the point source in the 1340-1710 \AA\ band.  The source is low
contrast and superposed on the bright galaxy core, which has a strong
radial brightness gradient.  We measured the excess flux over the
estimated background within a $\pm2.5$ pixel region centered on the
source.  We then extrapolated to total flux using the known point
source encircled energy function for the spectrograph, as given in the
STScI {\it STIS Instrument Handbook}.  The main uncertainty in this and
the other measurements reported here arises from the interpolation of
the background light to the nuclear position.  Error bars were
estimated from interpolations using maximum or minimum slopes for the
background light profile. 

We find that the nuclear source has a mean flux density $f_{\lambda}
\sim 1.6 \pm 0.4 \times 10^{-17}\, {\rm erg\ s}^{-1}\ {\rm cm}^{-2}\
{\rm \AA}^{-1}$ in the 1340-1710 \AA\ band, or $m_{\lambda}(1525\,
{\rm \AA}) = 20.9$ in the STMAG monochromatic system ($m_{\lambda} =
-2.5 \log f_{\lambda} -21.1$, where $f_{\lambda}$ is in units of ergs
s$^{-1}$ cm$^{-2}$ \AA$^{-1}$; Holtzman et al.\ 1995).  The energy
distribution of the source is roughly flat in $f_{\lambda}$ units over
the FUV band, and no emission or absorption features are apparent
(though such would be heavily smoothed by the wide effective
bandwidth).

The point source is also detectable in our optical-band STIS/CCD
spectrum, though with lower contrast.  It shows a smooth continuum, without
emission lines and also without the strong absorption lines
characteristic of the cool stars in the galaxy background light.  We
made an extraction of the 2-D spectral image corresponding to the
coverage of the WFPC2 F450W filter (3950-4870 \AA) and, using the method
described above, measured the
mean flux density of the nuclear source to be $f_{\lambda} \sim 6.4
\pm 1.3 \times 10^{-18}\, {\rm erg\ s}^{-1}\ {\rm cm}^{-2}\ {\rm \AA}^{-1}$ in
this band. 

Our optical spectrum does not extend completely through the regions
covered by the F555W or F606W broad-band imaging filters on the HST
cameras.  However, to provide a rough comparison, we extracted a mean
flux density for the nuclear source in the 400 \AA\ region centered on 5430
\AA\ of $f_{\lambda} \sim 6.7
\pm 2.2 \times 10^{-18}\, {\rm erg\ s}^{-1}\ {\rm cm}^{-2}\ {\rm \AA}^{-1}$.

There are long-exposure images of the center of NGC 1399 at four other
epochs in the HST archives.  Two of these have been discussed in the
literature.  Renzini et al.\ (1995) and Cappellari et al.\ (1999) had
scrutinized the 5 October 1993 FOC/UV image of NGC 1399 and found a
perfectly smooth light distribution.  There was no point source
present comparable to the one they detected on similar FOC/UV imaging
in NGC 4552.  Likewise, Stiavelli et al.\ (1993) analyzed the 7
November 1991 WFPC1/PC image of NGC 1399 in the V band, comparing it
to high resolution, ground-based R band images.  Although the central
pixel of the HST image showed a small excess, there was no evidence
for the expected wings of the point-spread function (PSF) at larger
radii.  They found distributed light in excess of an isothermal core
on both image sets, but this did not correspond in structure to that
expected from a point source.  Lauer et al.\ (1995) analyzed the same
WFPC1/PC image as Stiavelli et al.\ (1993) and again found no evidence
for a nuclear point source. Note that although both the 1991 and 1993
HST observations were made with the aberrated HST optics, which produced
large image wings, the core of the aberrated PSF is quite sharp (FWHM
$\sim 0.04\arcsec$) and would permit detection of a nuclear
point source. 

None of these earlier studies provided quantitative limits on the flux
from a nuclear point source, so we re-analyzed the archival images.
That included two additional WFPC2 observations made with the PC chip
in 1996 and 2000 (with full correction for the spherical aberration).
NGC 1399 has nearly circular isophotes, so we performed photometry on
the images with circular or annular apertures centered on the point of
light symmetry.  We interpolated the background surface brightness to
the nuclear position and measured the excess light in a central
aperture of between 2 and 5 pixel radius, depending on the image.  We
then extrapolated to total flux using the known encircled energy
functions for each camera and filter as given in the STScI handbooks.
Again, the dominant source of photometric error in this process is the
interpolation of the bright stellar background light to the galaxy
center; error bars or limits were estimated as previously described.
Results of the photometry are given in Table 1 as detections or
$3\sigma$ upper limits.  

Apart from the 1999 STIS observations, only in the case of the 2 June
1996 WFPC2 F450W image do we detect a point source.  A plot of the
central light profile from that image is shown in Figure 2. The
signature of a point source is clear, although its effects are
confined to the innermost 5 pixels ($r \la 0.06\arcsec$). 

In 1999, the UV/optical color of the source was
{\nobreak $m_{\lambda}(1525\, {\rm
\AA}) - m_{\lambda}(4480\, {\rm
\AA}) \sim -1.0$,}
corresponding to the energy distribution of a late B
star or a power-law index of $\alpha \sim 1.2$, where $f_{\nu} \sim
\nu^{-\alpha}$.  This is a softer spectrum than is typical of AGN but
is within the rather large dispersion observed in AGN continua (e.g.\
Risaliti \& Elvis 2004).  The extracted UV and optical spectra of the
source are each consistent, within the appreciable noise, with the
modest spectral slope ($f_{\lambda} \sim \lambda^{-0.8}$) implied by
this index.  

We do not believe extinction by dust has much effect on these
measurements at either optical or UV wavelengths.  Foreground Galactic
extinction is negligible, and no evidence of UV extinction was
detected on the spatial scales ($\sim 3$-60\arcsec) accessible to the
Ultraviolet Imaging Telescope or the spectra from the Hopkins
Ultraviolet Telescope (Ferguson et al.\ 1991; O'Connell et al. 1992;
O'Neil et al.\ 1996; Ohl et al.\ 1998; Marcum et al.\ 2001).  Our STIS
spectrum has a smooth, symmetrical spatial profile, with no evidence
of the dusty lanes or disks found in some E galaxies.

\section{The Variable AGN in NGC 1399}

In the last column of Table 1 we have converted the monochromatic flux
densities or limits to luminosities per decade ($\lambda L_{\lambda}$)
for easier multiband comparisons.  This is a heterogeneous collection
of measurements in different bands.  However, because of its
flat energy distribution ($\lambda L_{\lambda} \sim
\lambda^{0.2}$), if the nuclear source were non-variable we would
have obtained approximately the same numerical values at all epochs in
all of the optical bands.  The table provides good evidence that the
source varied by a factor of $\ga 3$ in brightness at both UV and
optical wavelengths over the period 1991-2000.  We cannot determine
the duty cycle of the source or decide if the activity is part of a
continuing flickering or is an unusually large, flare-like outburst.
If only a single event was involved, it began by mid-1996.  Brightest
values were measured in 1999, and it appears to have faded by $\sim
4\times$ in the 16 months between January 1999 and May 2000. 

Normal stellar phenomena in old stellar populations are too faint to
produce this kind of nuclear source (cf.\ Renzini et al.\ 1995).  The
absolute monochromatic magnitude of the source in 1999 was
$M_{\lambda}(1525\, {\rm \AA}) \sim -10.6$.  The brightest known
globular clusters in our Galaxy and in the gE galaxy M87, for
instance, are fainter than this, at $-8.9$ and $-9.6$ respectively, in
the far-UV (Sohn et al.\ 2004).  The integrated light of clusters
also, of course, cannot change more than that of the variable stars
within them.  The temporal behavior of the source is not consistent
with a supernova.  Novae and other UV transient sources have been
detected in long-term HST monitoring programs of M87 directed by J.
Biretta and W. Sparks (e.g.\ Sparks et al.\ 2000) and in a shorter
program by Shara et al.\ (2004) and Baltz et al.\ (2004).  However,
these are again considerably fainter than the NGC 1399 nuclear
source.  For instance, 11 novae in M87 were identified by Sohn et al.\
(2004), but they are all fainter than $M_{\lambda}(2500\, {\rm \AA}) =
-7$.  Most nova outbursts also do not persist for more than about 6
months, so multiple events would be required to explain the
observations.  Several kinds of hot massive stars (e.g.\ O3-O7 dwarfs,
OB supergiants, and Wolf-Rayet stars) have $m_{\lambda}(UV) -
m_{\lambda}(V) \la -4.0$ and can be both as bright as the nuclear
source and also exhibit variability.  However, there is no other
evidence for a very young stellar population in NGC 1399.  Finally,
the exact coincidence of the source with the nucleus of NGC 1399
together with the absence of comparable extra-nuclear sources on any
of the available data sets argues against a normal stellar
phenomenon.  

The only earlier evidence of nuclear activity in NGC 1399 was a
low-luminosity central radio source (Killeen, Bicknell \& Ekers 1988;
Sadler, Jenkins \& Kotanyi 1989).  This consists of a core and two
oppositely-directed jets feeding diffuse lobes.  The whole structure
is small and lies entirely within the optical body of the galaxy.
Corrected to our adopted distance for NGC 1399, the total radio
luminosity is $2.2 \times 10^{39}$ ergs s$^{-1}$, while the luminosity
in the unresolved core is only $6.6 \times 10^{37}$ ergs s$^{-1}$. 

X-ray observations with ROSAT and the {\it Chandra} X-Ray Observatory
have placed only upper limits on the X-ray luminosity of NGC 1399's
nucleus (Sulkanen \& Bregman 2001; Loewenstein et al.\ 2001).  The
{\it Chandra} limit is $L_{\rm X} < 9.7 \times 10^{38}$ ergs s$^{-1}$
in the 2-10 keV band (data taken on 18 January 2000).  There are,
however, over 200 extra-nuclear X-ray point sources detected, many
associated with globular clusters (Angelini, Loewenstein \& Mushotzky
2001).  These are presumably low-mass X-ray binaries.  There are, in
fact, so many of them that they interfere with obtaining better limits
on the nuclear X-ray brightness.  

The UV/optical nuclear point source we have found is almost certainly
the AGN responsible for the radio emission in NGC
1399.  Its luminosity at detection was $\lambda L_{\lambda}({\rm UV})
\sim 1.2
\times 10^{39}$ ergs s$^{-1}$. 

The activity in NGC 1399 is nearly identical in photometric properties
to the UV nuclear transient in NGC 4552 (Cappellari et al.\ 1999).
Given the fact that UV observations of E galaxies with HST have been
made only infrequently, the accidental discovery of two such episodes
in 6 years implies that high resolution UV monitoring is a productive
means of isolating nuclear activity in nearby galaxies.

\section{Discussion}

The radio, optical, and UV data discussed above demonstrate that NGC
1399 contains a variable AGN, but it is definitely an inconspicuous
one and falls in the LLAGN category.  Because of NGC 1399's proximity,
one can explore accretion of interstellar or intracluster gas by the
nucleus in ways not possible for more distant sources.  Loewenstein et
al.\ (2001) have used a kinematic estimate of the mass of the central
SMBH from Merrit \& Ferrarese (2001) together with the high resolution
{\it Chandra} X-ray images to estimate that the Bondi accretion rate
onto the SMBH from the surrounding diffuse, hot medium is $\sim 0.04\,
{\rm M}_{\odot}$ yr$^{-1}$.  The predicted bolometric luminosity,
assuming the standard 10\% efficiency for conversion of accreted
material to photons, is then $\sim 2 \times 10^{44}$ ergs s$^{-1}$.
But this is a factor of $10^5$ greater than the observed luminosity of
the nuclear source in any of the bands described above.  

This gross radiative inefficiency compared to the expectations of
simple inflow models is characteristic of LLAGNs.  It is possible that
the gas inflow rates have been seriously overestimated, and it is
usually difficult to obtain good observational constraints on the
flows at the necessary spatial scales.  More frequently, however, it
has been argued that the nuclear structure of LLAGNs may differ
fundamentally from those of more active systems, which are dominated
by optically thick accretion disks.  At very sub-Eddington accretion
rates, the structure of the gas flow near a SMBH may change in such a
way that little of the dissipation energy generated by the inflow
appears as electromagnetic radiation, instead being advected across
the event horizon or expelled in a wind or jet.  The resulting
efficiency of mass conversion to luminosity would be much below the
10\% typical of luminous AGN's.  Such structures are called RIAFs for
``radiatively inefficient accretion flows'' (e.g.\ Quataert 2003), and
there are several possible types of them (e.g.\ Narayan \& Yi 1994,
Fabian \& Rees 1995, Blandford \& Begelman 1999, Quataert \& Gruzinov
2000, Hawley \& Balbus 2002, and references therein), featuring a
variety of components such as spherical inflows, hot coronas, inner
tori, optically thin Keplerian disks, convection currents, and jets or
other outflows.  

A basic expectation of RIAF models, confirmed by numerical simulations
(Hawley \& Balbus 2002), is that the flow structure can be strongly
time-dependent and should respond sensitively to changes in the
overall mass accretion rate.  The resulting variations in the
broad-band spectrum are observable.  The source luminosity relative to
the accretion rate, the shape of the energy distribution, and the
variability with photon energy are key tests of the viability of the
various RIAF models (Quataert \& Narayan 1999, Ho 2002, DiMateo et
al.\ 2003). 

Vacuum-UV observations are important constraints on the models.  In
conventional dense accretion disk models, strong thermal continuum
emission from the inner disk generates a ``big blue bump'' at UV
wavelengths.  This can lie at the peak of the $\lambda L_{\lambda}$
energy distribution, and both X-ray and UV emission can vary rapidly
because of the small size of the emitting volume (e.g.\ Ptak et al.\
1998).  By contrast, RIAFs are thought to lack such dense disks at
small radii, and most models predict that emission will be suppressed
in the UV compared to the radio and X-ray regions.  UV photons are
produced over a large range of radii ranging up to the size of the
Bondi accretion radius.  This can be up to $\sim 50$ pc in typical
cases (DiMateo et al.\ 2003), implying only slow UV variability.  

In the case of NGC 1399, the RIAF models seem to have sufficient
flexibility to fit the UV/optical spectral slope we observed for the
NGC 1399 nucleus in 1999.  However, the relatively short variability
time scale ($\sim 1$-5 years) in the UV/optical bands compared with
the large Bondi radius of $\sim 35$ pc (Loewenstein et al.\ 2001)
appears to be incompatible with RIAF models.  Similarly, the
luminosity of the source is a difficulty for RIAF models. Loewenstein
et al.\ (2001) found that the X-ray upper limit for NGC 1399 is not
consistent with simple RIAF models given the estimated accretion rate,
being over 100$\times$ lower than predicted.  That would also be true
of the UV and optical luminosities given in Table 1.  The models
predict relatively large ratios of X-ray to UV luminosity in $\lambda
L_{\lambda}$ units, which are not supported by the values quoted
above.  Unfortunately, the UV and X-ray observations were not
contemporaneous (they are separated by a year), so cannot serve
as a good test of the X-ray/UV ratio. 

For M87, DiMateo et al.\ (2003) argue that the famous relativistic jet
may disrupt the RIAF inflow, suppressing radiation or yielding intermittent
accretion (see also Fabbiano et al.\ 2003 for the case of IC1459). The
jets themselves may carry off most of the accretion energy.  The radio
jets in NGC 1399 could well operate in a similar manner to suppress
the accretion luminosity there.

There are, of course, other sources of fuel for AGN's than the
surrounding diffuse gas.  In their interpretation of the NGC 4552 event,
Cappellari et al.\ (1999) favored a standard accretion disk model in
which partial stripping of a single stellar atmosphere by the central
SMBH (releasing $\sim 0.001 {\rm M}_{\odot}$) was responsible for a
flare-like energy release.  Now that a twin event has been detected in
NGC 1399, the likelihood of such stripping episodes should be
quantitatively assessed to determine whether this is still a viable
explanation.  Although the tidal destruction of an entire star is a
very rare occurrence (once every $10^4$-$10^5$ years, Rees 1990 and
Magorrian \& Tremaine 1999), partial stripping could be more common.
There is some evidence for strong X-ray flares consistent with tidal
disruption events in a small number of galaxies from the ROSAT
surveys (e.g.\ Gezari et al.\ 2003).  Other objects, e.g.\
planets, could also be the source of transient events.  

HST monitoring of nuclei with deep exposures at short wavelengths is
an excellent means of assessing LLAGN variability and probing fueling
mechanisms.  Very small levels of mass transfer in the nuclei can be
detected.  As noted above, X-ray detection in such situations can be
limited by stellar sources within the galaxy.  Neither NGC 1399 nor
4552 were actually optimal cases for UV point source detections
because of their abnormally bright FUV stellar backgrounds.  This
should be easier in galaxies with more normal UV colors, {\nobreak
$m_{\lambda}({\rm FUV}) - V
\ga 3$}.  There are several other instances of optical-band identifications
with HST of nuclear point sources in gE galaxies, including NGC 4278
(Carollo et al.\ 1997), IC 1459 (Fabbiano et al.\ 2003), and NGC 4374
(Bower et al.\ 2000) that would be good candidates for HST monitoring.

After we submitted this paper, Maoz et al.\ (2005) reported results
from an HST/ACS ``snapshot'' survey that confirms the value of high spatial
resolution UV monitoring.  They observed a sample of known UV-bright
LINERS in spiral and E galaxies at 2500 \AA\ at irregular intervals
over 12 months.  Most of the sources exhibit low-level variability,
and some with earlier observations show fractional changes comparable
to those of NGC 1399.   The survey demonstrates that AGN's are indeed
associated with many LINERS.  All the sample sources with detected radio
cores (as in NGC 1399) have variable UV nuclei.  Of the new
identifications, only NGC 4258 has a nuclear luminosity comparable to
or fainter than NGC 1399 or 4552. 

\section{Conclusion}

We detected a nuclear transient in the giant elliptical galaxy NGC
1399 on deep far-UV spectra taken with HST/STIS in January 1999.  The
source reached a maximum observed far-UV luminosity of $\sim 1.2 \times
10^{39}$ ergs s$^{-1}$ at that time.  Re-reduction of earlier HST
archival images showed that the source was detectable in an
optical-band image in 1996 but not in UV/optical images from 1991 or
1993.  It faded by a factor of $\sim 4\times$ by mid-2000.  We cannot
determine the duty cycle of the source. 

Ordinary stellar phenomena are not likely explanations of such
events.  Instead, the activity is almost certainly associated with the
low luminosity AGN responsible for the core/jet radio emission in NGC
1399.  The source is much fainter than expected from its Bondi
accretion rate (estimated from {\it Chandra} high resolution X-ray
images), even in the context of ``radiatively inefficient accretion
flow'' models, and its variability also appears inconsistent with such
models.  

NGC 1399 is of special interest because of its involvement with a
nearby cluster cooling flow (Rangarajan et al.\ 1995) and the growing
realization that AGN feedback probably plays a central role in
regulating gas accretion and therefore galaxy and SMBH growth in
clusters and in the early universe (e.g.\ Wu, Fabian, \& Nulsen 2000;
McNamara et al.\ 2005). 

The properties of the NGC 1399 outburst are remarkably similar to the
UV-bright nuclear transient discovered earlier with HST/FOC in NGC
4552 by Renzini et al.\ (1995).  There had been no earlier evidence
for activity at optical through X-ray wavelengths in either galaxy.

These accidental discoveries and the more recent detection of UV
variability in known LINERS by Maoz et al.\ (2005) demonstrate that
high resolution UV monitoring is a valuable means to study activity in
nearby LLAGNs.  Note that {\it Chandra} was able to place only an
upper limit to the X-ray emission from the NGC 1399 nucleus, partly
because of the presence of many X-ray bright binaries.  UV nuclear
emission will often present a greater contrast with the background
galaxy than will X-ray emission.  Especially when coupled with radio
core observations over long periods, UV monitoring will be a useful
discriminant between models for fueling activity in the vicinity of a
supermassive black hole.

\acknowledgements 

In the course of this work we utilized data from HST program numbers
2600, 3728, 5990, 7438, and 8214.  We are grateful to Charlie Wu for
his early participation in this program; to Richard Mushotzky for
sharing X-ray data on NGC 1399; and to John Hawley for comments on the
manuscript.  We thank the anonymous referee for comments that helped
improve the presentation.  This research was supported in part by
NASA LTSA grant NAG5-6403 and STScI grant GO-07438.


\newpage



\begin{deluxetable}{clcccr}
\tablewidth{0pt}
\tablecaption{ HST Observations of NGC 1399 Nuclear Source}
\tablehead{ \colhead{} & \colhead{} & \colhead{} & \colhead{$\bar\lambda$} 
          & \colhead{$f_{\lambda} \times 10^{18}$} 
          & \colhead{$\lambda L_{\lambda}$} \\

          \colhead{Date} & \colhead{Instrument} & \colhead{Filter}
		      & \colhead{ (\AA) } 
                      &  \colhead{ ( ${\rm erg\ s}^{-1}\ {\rm cm}^{-2}\ {\rm \AA}^{-1} $ )}
                      &  \colhead{ ( ${\rm erg\ s}^{-1}$ )  }
           }
\startdata
1991-11-07 & WFPC1/PC & F555W   & 5429 & $<3.5 $                & $<9.4 \times 10^{38}$ \\
1993-10-05 & FOC      & F175W   & 1730 & $<2.8 $                & $<2.4 \times 10^{38}$ \\
1996-06-02 & WFPC2/PC & F450W   & 4484 & $ \;\;\;\; 2.5 \pm 0.5$ & $ 5.5 \times 10^{38}$ \\
1999-01-10 & STIS     & G430L   & 4480 & $ \;\;\;\; 6.4 \pm 1.3$ & $ 1.4  \times 10^{39}$ \\
1999-01-10 & STIS     & G430L   & 5430 & $ \;\;\;\; 6.7 \pm 2.2$ & $ 1.8 \times 10^{39}$  \\
1999-01-19 & STIS     & G140L   & 1525 & $ \;\;\;\; 16.0 \pm 4.0$ & $ 1.2 \times 10^{39}$ \\
2000-05-18 & WFPC2/PC & F606W   & 5860 & $<1.5 $                 & $<4.3 \times 10^{38}$ \\
\enddata

\end{deluxetable}

\newpage



\begin{figure}[t]
\plotone{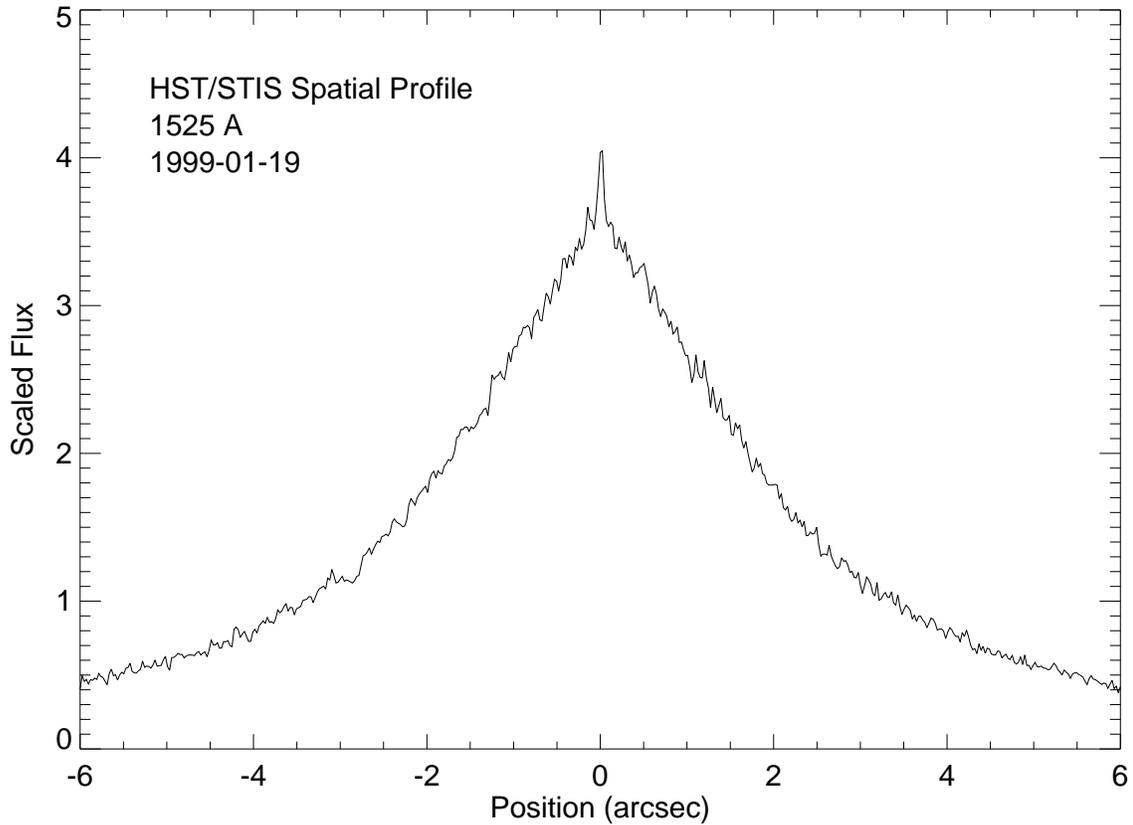}
\vspace{10pt}
\caption{ The spatial profile of UV light in the center of NGC 1399
from the STIS/FUV spectrum of January 1999.  Plotted is the mean surface
brightness per pixel (arbitrary flux units) for 0.024\arcsec\ high bins
within the 2\arcsec\ wide spectrograph slit as a function of distance along
the slit from the galaxy's light centroid.  The profile is an average
of the spectral image covering 1340-1710
\AA.  There is a distinct nuclear point source superposed on the
broader profile of background stellar light. 
 }
\end{figure}


\begin{figure}[ht]
\plotone{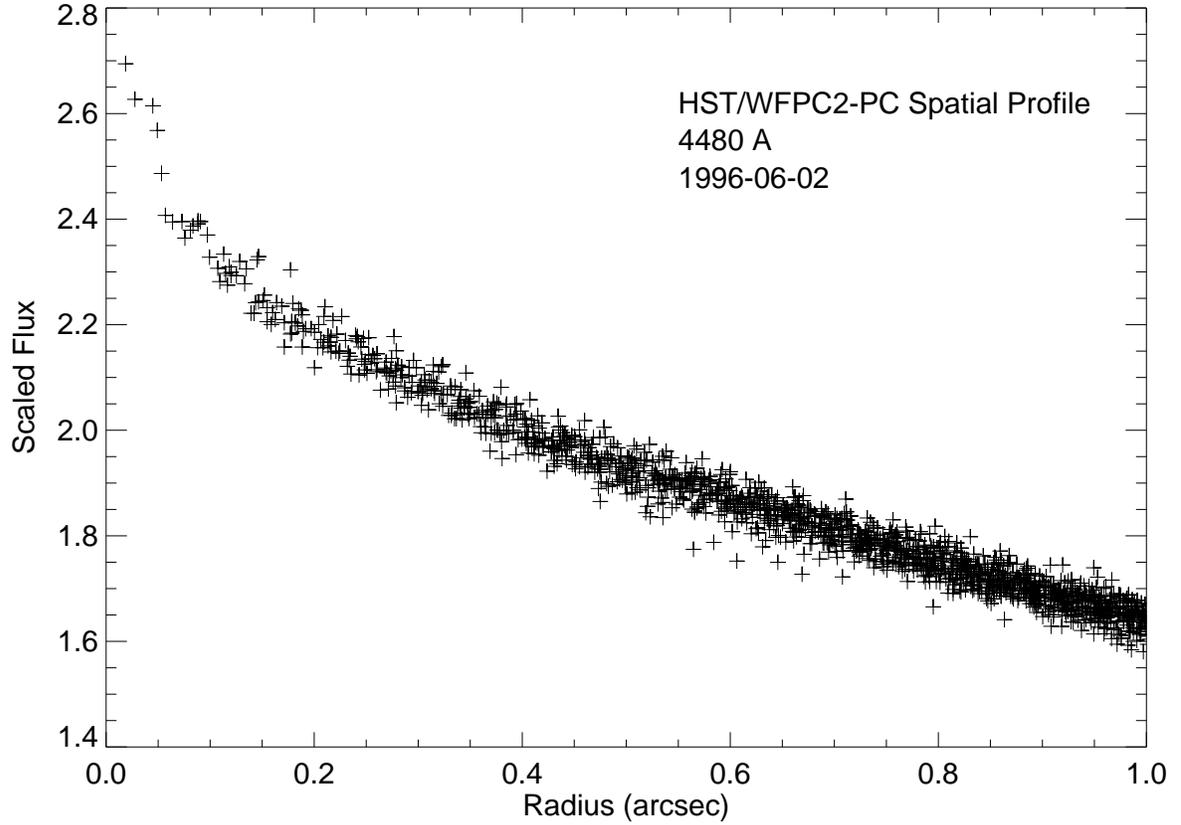}
\vspace{10pt}
\caption{
Central light distribution of NGC 1399 in the WFPC2-PC F450W image of
June 1996.  Plotted are the fluxes for individual pixels as a function of
distance from the light centroid.  The unit for the flux scale is $1.0
\times 10^{-18}\, {\rm erg\ s}^{-1}\ {\rm cm}^{-2}\ {\rm \AA}^{-1}$.
The pixel size is 0.045\arcsec.  Although the stellar background light rises
toward the center, there is a distinct nuclear point source superposed
on it.
  }
 \end{figure}
  
\end{document}